\documentclass[fleqn,10pt]{wlscirep}

\usepackage{multirow}
\usepackage{mathrsfs}

\title{Fringe pattern analysis using deep learning}

\author[1,2,3]{Shijie Feng}
\author[1,2,4]{Qian Chen}
\author[1,2]{Guohua Gu}
\author[1,2]{Tianyang Tao}
\author[1,2,3]{Liang Zhang}
\author[1,2,3]{Yan Hu}
\author[1,2,3]{Wei Yin}
\author[1,2,3,*]{Chao Zuo}

\affil[1]{School of Electronic and Optical Engineering, Nanjing University of Science and Technology, No. 200 Xiaolingwei Street, Nanjing, Jiangsu Province 210094, China}
\affil[2]{Jiangsu Key Laboratory of Spectral Imaging \& Intelligent Sense, Nanjing, Jiangsu Province 210094, China}
\affil[3]{Smart Computational Imaging Laboratory (SCILab), Nanjing University of Science and Technology, Nanjing, Jiangsu Province 210094, China}
\affil[4]{chenqian@njust.edu.cn}
\affil[*]{Corresponding author: zuochao@njust.edu.cn}

\begin{abstract}
In many optical metrology techniques, fringe pattern analysis is the central algorithm for recovering the underlying phase distribution from the recorded fringe patterns. Despite extensive research efforts for decades, how to extract the desired phase information, with the highest possible accuracy, from the minimum number of fringe patterns remains one of the most challenging open problems. Inspired by recent successes of deep learning techniques for computer vision and other applications, here, we demonstrate for the first time, to our knowledge, that the deep neural networks can be trained to perform fringe analysis, which substantially enhances the accuracy of phase demodulation from a single fringe pattern. The effectiveness of the proposed method is experimentally verified using carrier fringe patterns under the scenario of fringe projection profilometry. Experimental results demonstrate its superior performance in terms of high accuracy and edge-preserving over two representative single-frame techniques: Fourier transform profilometry and Windowed Fourier profilometry.
\end{abstract}
\begin{document}

\flushbottom
\maketitle
\thispagestyle{empty}

\section*{Introduction}

\noindent

Optical measurement techniques such as holographic interferometry \cite{kreis}, electronic speckle pattern interferometry \cite{Rastogi}, and fringe projection profilometry \cite{gorthi2010fringe} are quite popular for non-contact measurements in many areas of science and engineering, and have been extensively applied for measuring various physical quantities like displacement, strain, surface profile, refractive index, etc. In all these techniques, the information about the measured physical quantity is stored in the phase of a two-dimensional fringe pattern. The accuracy of measurements carried out by these optical techniques is thus fundamentally dependent on the accuracy with which the underlying phase distribution of the recorded fringe patterns is demodulated.

Over the past few decades, tremendous efforts have been devoted to developing various techniques for fringe analysis, and they can be broadly classified into two categories: (1) phase-shifting (PS) methods which require multiple fringe patterns to extract phase information \cite{Zuo_ps_review} and (2) spatial phase demodulation methods which allow phase retrieval from a single fringe pattern, such as the Fourier transform (FT) \cite{RN365}, windowed Fourier transform (WFT) \cite{RN38}, and wavelet transform (WT) methods. Compared to spatial phase demodulation methods, the multiple-shot phase-shifting techniques are generally more robust and can achieve pixel-wise phase measurement with higher resolution and accuracy. Furthermore, the phase-shifting measurements are quite insensitive to non-uniform background intensity and fringe modulation. Nevertheless, due to their multi-shot nature, these methods are difficult to apply for dynamic measurements and are more susceptible to external disturbance and vibrations. Thus, for many applications, phase extraction from a single fringe pattern is desired, which falls under the purview of spatial fringe analysis. In contrast to phase-shifting techniques where the phase map is demodulated on a pixel-by-pixel basis, the phase estimation at a pixel in spatial methods is influenced by its neighborhoods or even all pixels in the fringe pattern, which provides better tolerance to noise at the expense of poor performance around discontinuities and isolated regions in the phase map \cite{Leihuang,zonghua}. Here, we demonstrate experimentally for the first time, to our knowledge, that the use of deep neural network can substantially enhance the accuracy of phase demodulation from a single fringe pattern. Deep learning is a powerful machine learning technique that employs artificial neural networks with multiple layers of increasingly richer functionality and has shown great success in numerous applications for which data are abundant \cite{RN399,RN400}.
\section*{Principle}

In our method, the network configuration is inspired by the basic process of most phase demodulation techniques, which are briefly recalled as follows. The mathematical form of a typical fringe pattern can be represented as
\begin{equation}
I(x,y) = A(x,y) + B(x,y)\cos \phi (x,y) \label{Eq1}
\end{equation}
where $I(x,y)$ is the intensity of the fringe pattern, $A(x,y)$ is the background intensity, $B(x,y)$ is the fringe amplitude, $\phi(x,y)$ is the desired phase distribution. Also, x and y refer to the spatial coordinates or pixels along the horizontal and vertical directions. In most phase demodulation techniques, the background intensity $A(x,y)$ is regarded as a
disturbance term and should be removed from the total intensity. Then a wrapped phase map is recovered from an inverse trigonometric function whose argument is a ratio for which the numerator characterizes the phase sine [$sin\phi(x,y)$] and the denominator characterizes the phase cosine [$cos\phi(x,y)$].

\begin{equation}
\phi (x,y) =  \arctan \frac{{M(x,y)}} {{D(x,y)}} = \arctan \frac{{cB(x,y)\sin \phi (x,y)}} {{cB(x,y)\cos \phi (x,y)}} \label{Eq2}
\end{equation}
where c is a constant depending on the phase demodulation algorithm used (e.g., in FTP $c = 0.5$, in $N$-step PS $c = N/2$), $M(x,y)$ and $D(x,y)$ represent the shorthands for the numerator and denominator terms, respectively. Note that the signs of $M(x,y)$ and $D(x,y)$ can be further used to uniquely define a quadrant for each calculation of $\phi(x,y)$. With the 4-quadrant phasor space, the phase values at each point can be determined modulo $2\pi$.

\begin{figure}[htbp]
\centering
\includegraphics[width=0.8\linewidth]{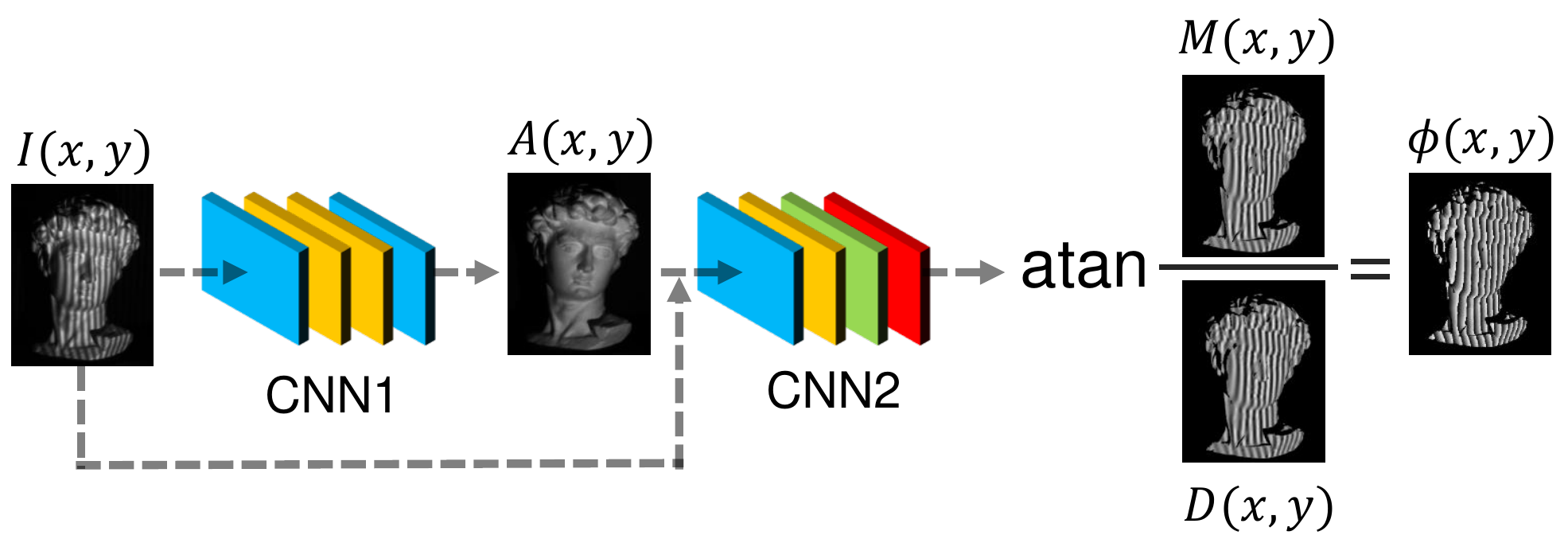}
\caption{Flowchart of the proposed method.}
\label{Fig1}
\end{figure}

In order to emulate the above process, two different convolutional neural networks (CNN) are constructed and are connected cascadedly according to Fig. \ref{Fig1}. The first convolutional neural network (CNN1) uses the raw fringe pattern $I(x,y)$  as input and estimates the background intensity $A(x,y)$ of the fringe pattern. With the estimated background image $A(x,y)$ and the original fringe image $I(x,y)$, the second convolutional neural network (CNN2) is trained to predict the numerator $M(x,y)$ and the denominator $D(x,y)$ of the arctangent function, which are fed into the subsequent arctangent function (Eq. \ref{Eq2}) to obtain the final phase distribution $\phi (x,y)$.

To generate the ground truth data used to train the two convolutional neural networks, phase retrieval is achieved by using a $N$-step phase-shifting method. The corresponding $N$ phase-shifted fringe patterns acquired can be represented as

\begin{equation}
{I_n}(x,y) = A(x,y) + B(x,y)\cos \left[ {\phi (x,y) - {\delta _n}} \right] \label{Eq3}
\end{equation}
where the index $n = 0,1,...,N-1$, and ${\delta _n}$ is the phase shift which equals $\frac{{2\pi n}}{N}$. With the orthogonality of trigonometric functions, the background intensity can be obtained by

\begin{equation}
A(x,y) = \frac{1}{N}\sum\limits_{n = 0}^{N-1} {{I_n}(x,y)} \label{Eq4}
\end{equation}
With the least square method, the phase can be calculated by
\begin{equation}
\phi (x,y) =  \arctan \frac{{\sum\nolimits_{n = 0}^{N-1} {{I_n}(x,y)\sin {\delta _n}} }} {{\sum\nolimits_{n = 0}^{N-1} {{I_n}(x,y)\cos {\delta _n}} }} \label{Eq5}
\end{equation}
Thus, the numerator and the denominator of the arctangent function in Eq. \ref{Eq2} can be expressed as

\begin{equation}
M(x,y){\rm{ = }}\sum\nolimits_{n = 1}^{N-1} {{I_n}(x,y)\sin {\delta _n}}  =  \frac{N}{2}B(x,y)\sin \phi (x,y) \label{Eq6}
\end{equation}

\begin{equation}
D(x,y){\rm{ = }}\sum\nolimits_{n = 0}^{N-1} {{I_n}(x,y)\cos {\delta _n}}  = \frac{N}{2}B(x,y)\cos \phi (x,y) \label{Eq7}
\end{equation}
The simplified expressions show that the numerator $M(x,y)$ and the denominator $D(x,y)$ are closely related to the original fringe pattern in Eq. \ref{Eq1} through a quasi-linear relationship with the background image $A(x,y)$. Thus, $M(x,y)$ and $D(x,y)$ can be learned by deep neural networks with ease with the knowledge of $A(x,y)$, which justifies our network design once again. It should be noted that the simple input - output network structure [linking fringe pattern $I(x,y)$ to phase $\phi(x,y)$ directly] performs poorly in our case since it is difficult to follow the phase wraps ($2\pi$ jumps) in the phase map precisely. Therefore, instead of estimating the phase directly, our deep neural networks are trained to predict the intermediate results, i.e., the numerator and the denominator of the arctangent function in Eq. \ref{Eq2} to obtain a better phase estimate.

\begin{figure}[htbp]
\centering
\includegraphics[width=0.8\linewidth]{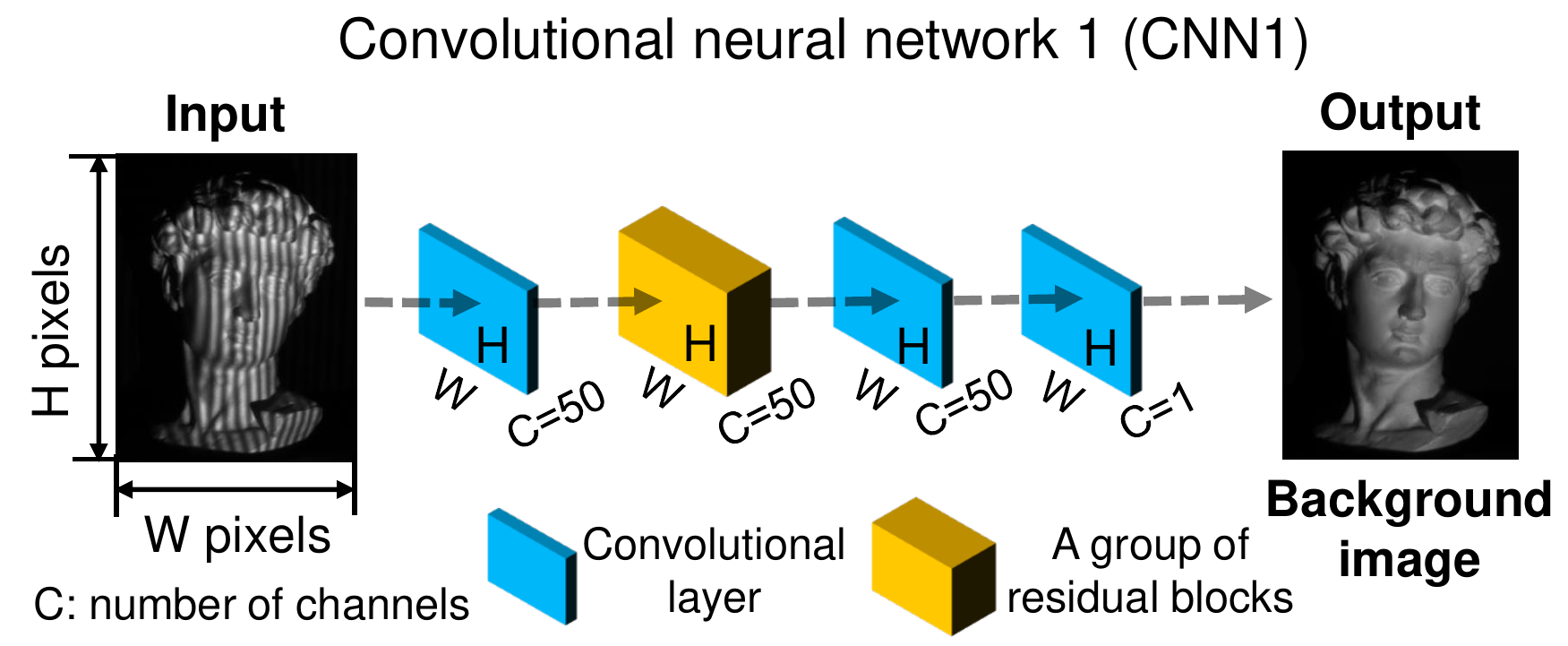}
\caption{Schematic of our first convolutional network (CNN1).}
\label{Fig2}
\end{figure}

\begin{figure}[htbp]
\centering
\includegraphics[width=0.8\linewidth]{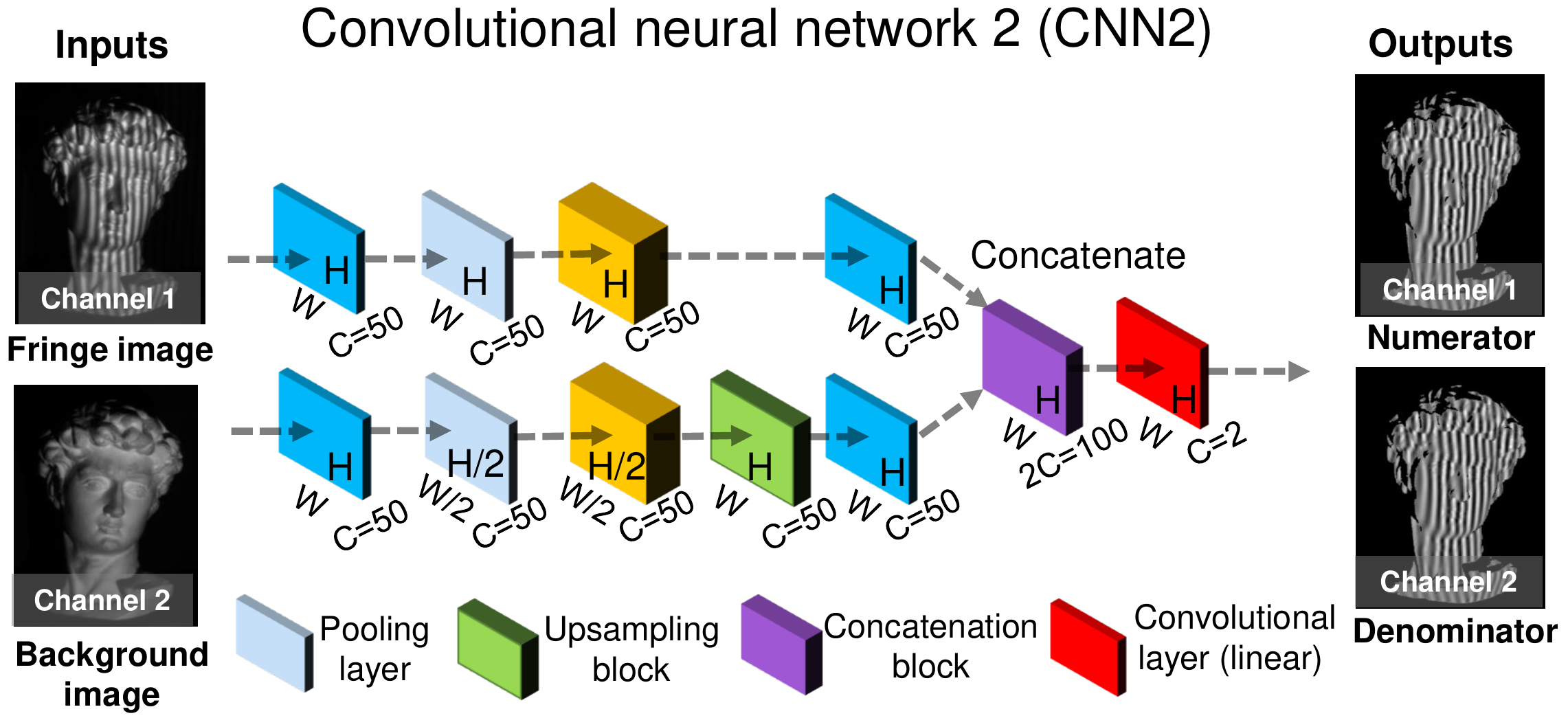}
\caption{Schematic of our second convolutional network (CNN2).}
\label{Fig3}
\end{figure}

To further reveal the internal structure of the two networks, the diagrams of the two convolutional neural networks are shown in Figs. \ref{Fig2} and \ref{Fig3}, respectively. The labeled dimension of the layers or the blocks shows the size of their output data. The input of CNN1 is a raw fringe pattern with $W\times H$  pixels. It is then successively processed by a convolutional layer, a group of residual blocks (containing four residual blocks) and two convolutional layers. The last layer estimates the gray values of the background image. With the predicted background intensity and the raw fringe pattern, the second convolutional neural network (CNN2) shown in Fig. \ref{Fig3} calculates the final phase distribution. In CNN2, the input images are down-sampled by $ \times 1$ and $ \times 2$ in two different paths. In the second path, the data is first downsamped for a high-level perception and then upsampled to match the original dimensions. With the two-scale data flow paths, the network can perceive more surface details for the numerator $M(x,y)$ and the denominator $D(x,y)$. We provide more details about the architectures of our networks in Supplement 1, Section 3.

\section*{Experiment}

The performance of the proposed approach was demonstrated under the scenario of fringe projection profilometry. The experiment consisted of two steps: training the testing.
In order to obtain the ground-truth training data, 12-step phase-shifting patterns with spatial frequency $f=160$ were created and projected by our projector (DLP 4500, TI) onto various objects. The fringe images were captured by a CMOS camera (8-bit pixel depth, resolution $640 \times 480$) simultaneously. Our training data set consists of 960 such fringe patterns and their corresponding ground truth data obtained by 12-step phase-shifting method (see Supplement 1, Sections 1 and 2 for details about the optical set-up and the collection of training data). Since one of the inputs of CNN2 is the output of CNN1, CNN1 was trained first and CNN2 was trained with the predicted background intensities and captured fringe patterns. These two neural networks were implemented using TensorFlow framework (Google) and was computed on a GTX Titan graphics card (NVIDIA).To monitor during training the accuracy of the neural networks on data that they have never seen before, we created a validation set by setting apart 150 fringe images from the original training data. Additional details on the training of our networks are provided in Supplement 1, Section 3.

\begin{figure}[htbp]
\centering
\includegraphics[width=0.8\linewidth]{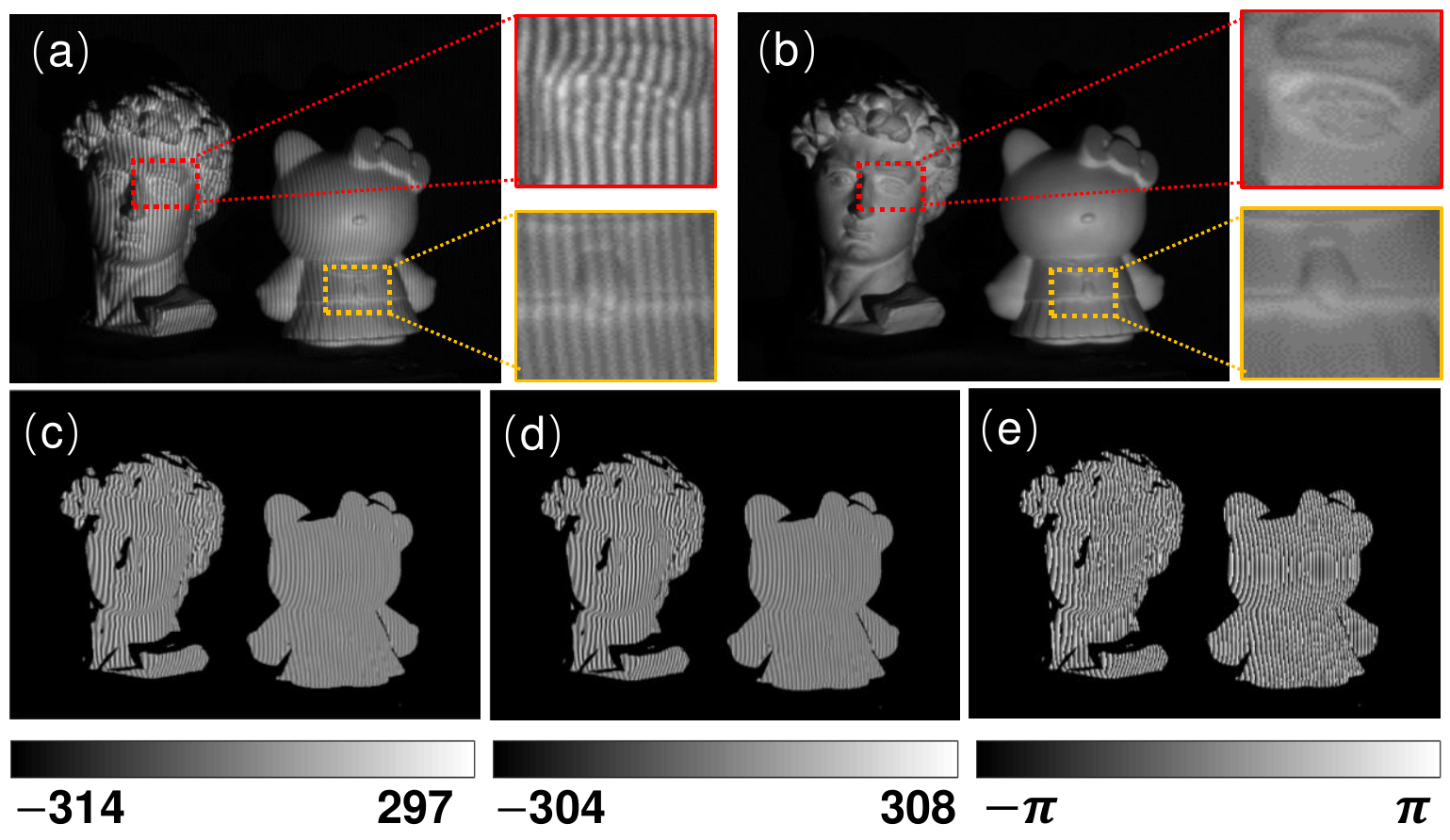}
\caption{Testing using the trained networks. (a) Input fringe image; (b) background image predicted by CNN1; (c) and (d) numerator and denominator estimated by CNN2; (e) phase calculated from (c) and (d).}
\label{Fig4}
\end{figure}

To test the trained neural networks versus conventional approaches, we measured a scene containing two isolated plaster models as shown Fig. \ref{Fig4}(a). Compared with the right model, the left one has a more complex surface, e.g., the curly hair and the high-bridged nose. It is noted that this scenario was never seen by our neural networks during the training stage. The trained CNN1 using the Fig. \ref{Fig4}(a) as an input predicted a background intensity as shown in Fig. \ref{Fig4}(b). From the enlarged views, we can see that the fringes have been removed completely through the deep neural network. Then, the trained CNN2 took the fringe pattern and the predicted background intensity as inputs and estimated the numerator $M(x,y)$ and the denominator $D(x,y)$, whose results are shown in Figs. \ref{Fig4}(c) and (d), respectively. The phase was calculated by Eq. \ref{Eq2} and is shown in Fig. \ref{Fig4}(e). To demonstrate the accuracy of determined phase, we unwrapped it by multi-frequency temporal phase unwrapping \cite{RN55} and calculated the phase error against the result recovered by 12-step phase-shifting method. In addition, two classic single-frame fringe analysis methods: FT \cite{RN365} and WFT \cite{RN38} were also implemented for comparison. Note that the parameters of these two algorithms were manually adjusted in order to obtain optimal results, while our method is completely non-parametric.
\begin{figure}[htbp]
\centering
\includegraphics[width=0.8\linewidth]{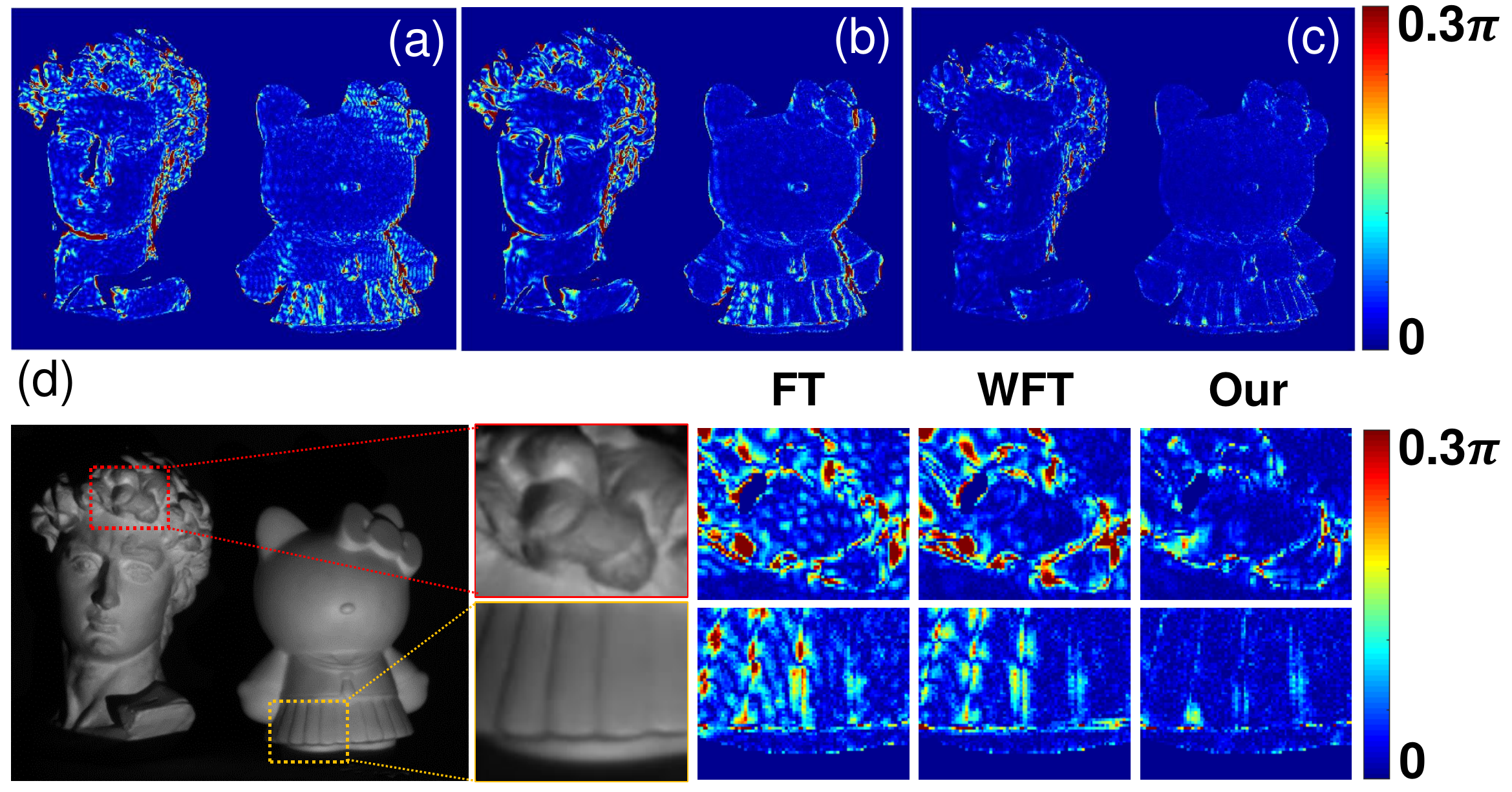}
\caption{Distribution of the phase error. (a) FT; (b) WFT; (C) our method; (d) comparison of details.}
\label{Fig5}
\end{figure}

Figures \ref{Fig5}(a)-(c) show the overall absolute phase error of these approaches. The result of FT shows the most prominent phase distortion as well as the largest mean absolute error (MAE) of 0.21 rad. WFT performed better than FT, with less significant errors especially for the right model (MAE 0.18 rad). Among these approaches, the proposed deep learning based method demonstrates the least errors for both models (MAE 0.08 rad). To illustrate the error distributions more clearly, two complex areas are enlarged and compared in Fig. \ref{Fig5}(d): the hair of the left model and the skirt of the right one. From Fig. \ref{Fig5}(d), obvious errors can be observed on the results of FT and WFT, which are mainly concentrated in the boundaries or abrupt depth-changing regions. By contrast, our approach greatly reduced the phase distortion, demonstrating its significantly improved performance in measuring objects with discontinuities and isolated complex surfaces.

\begin{figure}[htbp]
\centering
\includegraphics[width=0.8\linewidth]{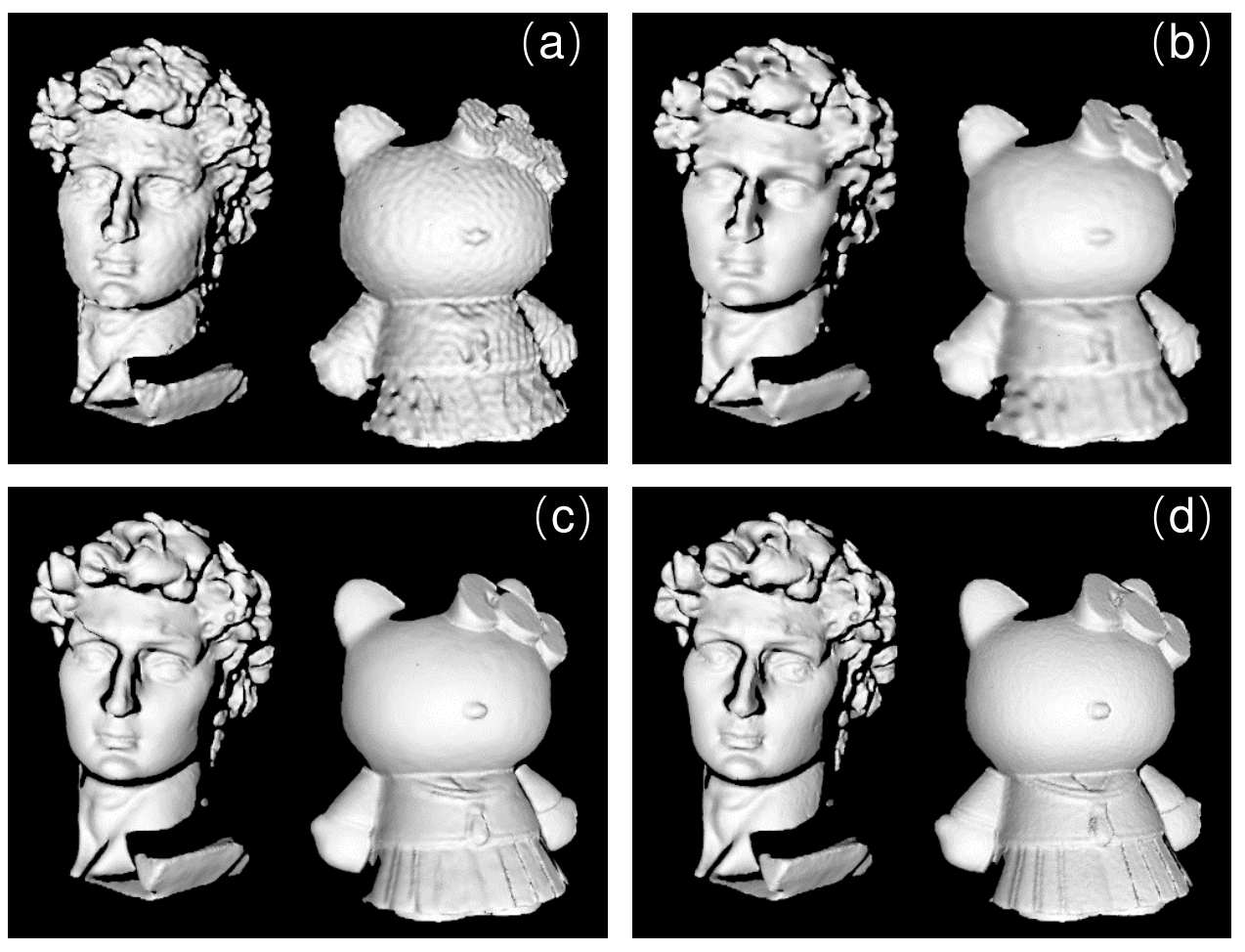}
\caption{3-D reconstructions. (a) FT; (b) WFT; (c) our method; (d) ground truth.}
\label{Fig6}
\end{figure}

For a more intuitive comparison, we converted the unwrapped phase into 3-D rendered geometries through stereo triangulation \cite{Zuo_OE}, as shown in Fig. \ref{Fig6}. From the visual inspection of Fig. \ref{Fig6}(a), the reconstructed result from FT features many grainy distortions, which are mainly due to the inevitable spectra leakage and overlapping in the frequency domain. Compared with FT, the WFT reconstructed the objects with more smooth surfaces but failed to preserve the surface details, e.g., the eyes of the left model and the wrinkles of the skirt of the right model, as can be seen in Fig. \ref{Fig6}(b). Among these reconstructions, the deep learning based approach yielded the highest-quality 3-D reconstruction [Fig. \ref{Fig6}(c)], which visually almost reproduced the ground truth data [Fig. \ref{Fig6}(d)] where 12-step phase-shifted fringe patterns were used. Besides, we also found the deep learning based fringe analysis is less sensitive to the frequency variation of captured fringes compared with FT and WFT. With a fringe pattern of a relatively low frequency (e.g., $f=60$), our method can still determine the phase distribution robustly. More details are provided in Supplement 1, Section 4.

\begin{figure}[htbp]
\centering
\includegraphics[width=0.8\linewidth]{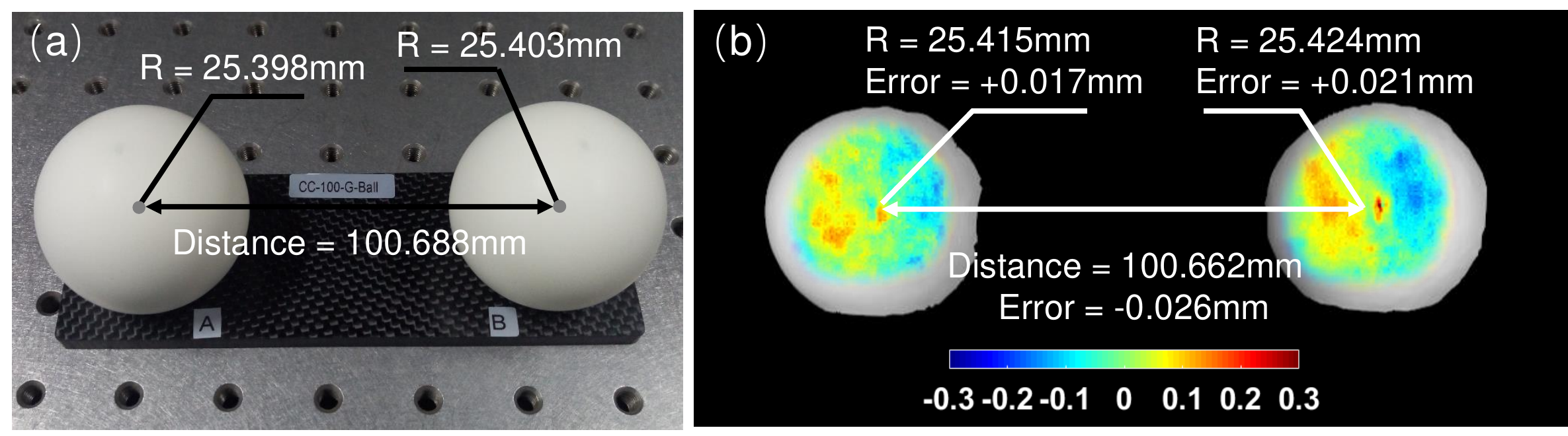}
\caption{ Quantitative analysis of the reconstruction accuracy of the proposed method. (a) A pair of standard spheres; (b) the reconstruction result.}
\label{Fig7}
\end{figure}

Finally, to quantitatively determine the accuracy of learned phase after converting to the desired physical quantity, i.e., 3-D shape of the object, we measured a pair of standard ceramic spheres whose shapes have been calibrated based on a coordinate measurement machine. Figure \ref{Fig7} (a) shows the tested ceramic spheres. The radii of them are 25.398mm and 25.403 mm, respectively, and their center-to-center distance is 100.688 mm.  We calculated the 3-D point cloud from the phase obtained by the proposed method and then fitted the 3-D points into the sphere model. The reconstructed result is shown in Fig. \ref{Fig7}(b), where the 'jet' colormap is used to represent data values of reconstruction errors. The radii of reconstructed spheres are 25.415 mm and 25.424 mm, with the deviations of 17$\mu m$ and 21$\mu m$, respectively. The measured center-to-center distance is 100.662mm, whose error is -26$\mu m$ . The tiny area having a relatively large error ($\approx 0.3{\rm{mm}}$) near the center of the right sphere is caused by the pixel saturation, since this area is so smooth that the specular components dominates the captured intensity. As the measured dimensions are very close to the ground truth, this experiment validates that the fringe analysis through deep learning not only provides reliable phase information using only single fringe pattern, but also facilitates high-accuracy single-shot 3-D measurements.

\section*{Conclusion}
We have demonstrated how the deep learning significantly improves the accuracy of phase demodulation from a single fringe pattern. Compared with existing single-frame approaches, this deep learning-based technique provides a new framework in fringe analysis by rapidly predicting the background image and estimating the numerator and the denominator for the arctangent function, resulting in high-accuracy edge-preserving phase reconstruction without the need for user-specified parameters. The effectiveness of the proposed method has been verified using carrier fringe patterns under the scenario of fringe projection profilometry. We believe that, after appropriate training with different types of data, the proposed network framework or its derivation should also be applicable to other forms of fringe patterns (e.g., exponential phase fringe patterns or closed fringe patterns) and other phase measurement techniques for immense promising applications.

\section*{Funding Information}
National Key R\&D Program of China (2017YFF0106403); National Natural Science Fund of China (61722506, 61705105, 11574152); China Postdoctoral Science Foundation (2017M621747); Jiangsu Planned Projects for Postdoctoral Research Funds (1701038A).

%\bigskip \noindent See \href{link}{Supplement 1} for supporting content.

%\section*{References}

% Bibliography
\bibliography{sample}

%\bibliographyfullrefs{sample}

\end{document}